\documentstyle[12pt,preprint,psfig,natbib]{aastex}  

\newcommand{\km}{km s$^{-1}$}
\newcommand{\vy}[2]{#1_{\scriptscriptstyle #2}}
\newcommand{\Ly}{Ly$\alpha$}

\def\gtorder{\mathrel{\raise.3ex\hbox{$>$}\mkern-14mu
             \lower0.6ex\hbox{$\sim$}}}
\def\ltorder{\mathrel{\raise.3ex\hbox{$<$}\mkern-14mu
             \lower0.6ex\hbox{$\sim$}}}
\def\proptwid{\mathrel{\raise.3ex\hbox{$\propto$}\mkern-14mu
             \lower0.6ex\hbox{$\sim$}}}
\def\0946{PG~0946+301}

\def\arcsec{\ifmmode '' \else $''$\fi}

\def\arcsecpoint{\ifmmode ''\!. \else $''\!.$\fi}

\def\kms{\ifmmode {\rm km\ s}^{-1} \else km s$^{-1}$\fi}
\def\Msun{\ifmmode {\rm M}_{\odot} \else M$_{\odot}$\fi}
\def\Lsun{\ifmmode {\rm L}_{\odot} \else L$_{\odot}$\fi}
\def\Zsun{\ifmmode {\rm Z}_{\odot} \else Z$_{\odot}$\fi}

\def\ergscm2{ergs\,s$^{-1}$\,cm$^{-2}$}
\def\icm3{{\rm cm}^{-3}}
\def\icm2{{\rm cm}^{-2}}
\def\qo{\ifmmode q_{\rm o} \else $q_{\rm o}$\fi}
\def\Ho{\ifmmode H_{\rm o} \else $H_{\rm o}$\fi}
\def\ho{\ifmmode h_{\rm o} \else $h_{\rm o}$\fi}

\def\vFWHM{\ifmmode v_{\mbox{\tiny FWHM}} \else
            $v_{\mbox{\tiny FWHM}}$\fi}
\def\CCF{\ifmmode F_{\it CCF} \else $F_{\it CCF}$\fi}
\def\ACF{\ifmmode F_{\it ACF} \else $F_{\it ACF}$\fi}
\def\Halpha{\ifmmode {\rm H}\alpha \else H$\alpha$\fi}
\def\Hbeta{\ifmmode {\rm H}\beta \else H$\beta$\fi}
\def\Hgamma{\ifmmode {\rm H}\gamma \else H$\gamma$\fi}
\def\Hdelta{\ifmmode {\rm H}\delta \else H$\delta$\fi}
\def\Lya{\ifmmode {\rm Ly}\alpha \else Ly$\alpha$\fi}
\def\Lyb{\ifmmode {\rm Ly}\beta \else Ly$\beta$\fi}
\def\Lyg{\ifmmode {\rm Ly}\beta \else Ly$\gamma$\fi}
\def\hi{H\,{\sc i}}

\def\hei{He\,{\sc i}}
\def\heii{He\,{\sc ii}}

\def\cii{C\,{\sc ii}}
\def\ciii{\ifmmode {\rm C}\,{\sc iii} \else C\,{\sc iii}\fi}
\def\civ{\ifmmode {\rm C}\,{\sc iv} \else C\,{\sc iv}\fi}

\def\nii{N\,{\sc ii}}
\def\niii{N\,{\sc iii}}
\def\niv{N\,{\sc iv}}
\def\nv{N\,{\sc v}}

\def\oiii{O\,{\sc iii}}
\def\o5007{[O\,{\sc iii}]\,$\lambda5007$}
\def\oiv{O\,{\sc iv}}
\def\ov{O\,{\sc v}}
\def\ovi{O\,{\sc vi}}

\def\neiv{Ne\,{\sc iv}}
\def\nev{Ne\,{\sc v}}
\def\nevi{Ne\,{\sc vi}}
\def\neviii{Ne\,{\sc viii}}

\def\mgx{Mg\,{\sc x}}
\def\siiv{Si\,{\sc iv}}
\def\siIV{Si\,{\sc iv}}
\def\siIII{Si\,{\sc iii}}
\def\siII{Si\,{\sc ii}}

\def\siii{S\,{\sc iii}}
\def\siv{S\,{\sc iv}}
\def\sv{S\,{\sc v}}
\def\svi{S\,{\sc vi}}

\def\ariv{Ar\,{\sc iv}}
\def\arv{Ar\,{\sc v}}
\def\arvi{Ar\,{\sc vi}}
\def\arvii{Ar\,{\sc vii}}

\def\feiii{Fe\,{\sc iii}}
\def\feiv{Fe\,{\sc iv}}

\def\piv{P\,{\sc iv}}
\def\pv{P\,{\sc v}}

\def\o{\o}
\begin{document}

\title{HST STIS OBSERVATIONS OF PG 0946+301: \\
 THE HIGHEST QUALITY UV SPECTRUM OF A BALQSO}

\author{
Nahum Arav\altaffilmark{1,2}, 
Martijn de~Kool\altaffilmark{3}, 
Kirk T. Korista\altaffilmark{4},
D. Michael Crenshaw\altaffilmark{5,6},
Wil van Breugel\altaffilmark{7},
Michael Brotherton\altaffilmark{8},
Richard F. Green\altaffilmark{8},
Max Pettini\altaffilmark{9},
Bev Wills\altaffilmark{10},
Wim de Vries\altaffilmark{7},
Bob Becker\altaffilmark{7,2},
W. N. Brandt\altaffilmark{11},
Paul Green\altaffilmark{14},
Vesa T. Junkkarinen\altaffilmark{15},
Anuradha Koratkar\altaffilmark{16},
Ari Laor\altaffilmark{17},
Sally A. Laurent-Muehleisen\altaffilmark{2},
Smita Mathur\altaffilmark{14},
Norman Murray\altaffilmark{18},
}


\altaffiltext{1}{Astronomy Department, UC Berkeley, Berkeley, 
CA 94720, I:arav@astron.Berkeley.EDU}
\altaffiltext{2}{Physics Department, University of California, Davis, CA 95616}
\altaffiltext{3}{Research School of Astronomy and Astrophysics, ANU ACT,
 Australia}
\altaffiltext{4}{Western Michigan Univ.,Dept. of Physics,1120 Everett Tower, 
Kalamazoo, MI  49008}
\altaffiltext{5}{Catholic University of America and Laboratory for Astronomy and
Solar Physics, NASA's Goddard Space Flight Center, Code 681,
Greenbelt, MD  20771}
\altaffiltext{6}{GSFC}
\altaffiltext{7}{IGPP, LLNL, L-413, P.O. Box 808 Livermore, CA 94550}
\altaffiltext{8}{Kitt Peak National Observatory,  950 North Cherry Avenue, P. O. Box
26732, Tuscon, AZ 85726}
\altaffiltext{9}{Institute of Astronomy, Cambridge, England}
\altaffiltext{10}{University of Texas}
\altaffiltext{11}{Department of Astronomy and Astrophysics, The Pennsylvania State University,
525 Davey Lab, University Park, PA 16802}
\altaffiltext{14}{Smithsonian Astrophysical Observatory}
\altaffiltext{15}{Center for Astrophysics and Space Sciences,
UCSD, 9500 Gilman Dr. La Jolla CA 92093}
\altaffiltext{16}{Space Telescope Science Institute}
\altaffiltext{17}{Technion, Israel}
\altaffiltext{18}{Canadian Institute for Theoretical Astrophysics,
University of Toronto}

\maketitle

\clearpage


\section*{ABSTRACT}
We describe deep (40 orbits) HST/STIS observations of the BALQSO PG
0946+301 and make them available to the community.  These observations
are the major part of a multi-wavelength campaign on this object aimed
at determining the ionization equilibrium and abundances (IEA) in
broad absorption line (BAL) QSOs. We present simple template fits to
the entire data set, which yield firm identifications for more than
two dozen BALs from 18 ions and give lower limits for the ionic column
densities.  We find that the outflow's metalicity is consistent with
being solar, while the abundance ratio of phosphorus to other metals
is at least ten times solar.  These findings are based on diagnostics
that are not sensitive to saturation and partial covering effects in
the BALs, which considerably weakened previous claims for enhanced
metalicity.  Ample evidence for these effects is seen in the spectrum.
We also discuss several options for extracting tighter IEA constraints
in future analyses, and present the significant temporal changes which
are detected between these spectra and those taken by the HST/FOS in
1992.

{\it Subject headings:} quasars: absorption lines --- quasars: individual (PG 0946+301)


\section{INTRODUCTION}

Broad Absorption Line (BAL) QSOs are a spectacular manifestation of AGN
outflows.  BALs are associated with prominent resonance lines such as
\civ~$\lambda$1549, \siiv~$\lambda$1397, \nv~$\lambda$1240, and \Ly\
$\lambda$1215.  They appear in about 10\% of all quasars (Foltz et al.\
1990) with typical velocity widths of $\sim10,000$ \km\ (Weymann,
Turnshek, \& Christiansen 1985; Turnshek 1988) and terminal velocities
of up to 50,000 \km.  The small percentage of BALQSOs among quasars is
generally interpreted as an orientation effect, and it is probable that
the majority of quasars and other types of AGN harbor intrinsic
outflows (Weymann et al.\ 1991).  

Establishing the ionization equilibrium and abundances (IEA) of the
BAL material is a fundamental issue in quasar studies (Weymann,
Turnshek, \& Christiansen 1985; Wampler, Chugai \& Petitjean 1995;
Hamann 1996; Korista et al.\ 1996; Arav et al.\ 2001; de~Kool et
al. 2001).  Furthermore, determining the IEA is crucial for
understanding the dynamics of the flows, especially radiative
acceleration scenarios that are strongly coupled to the ionization
equilibrium (Arav, Li \& Begelman 1994; de~Kool \& Begelman 1995;
Murray et al 1995; Arav 1996; Proga, Stone \& Kallman 2000).  Finally,
the mass flux and kinetic luminosity associated with the flow cannot
be constrained without a reliable IEA determination.

Careful spectroscopic analysis has made it apparent that the BALs are
often saturated while not black (Arav 1997; Telfer et al.\ 1998;
Churchill et al.\ 1999; Arav et al.\ 1999b; de~Kool et al.\ 2001).
Support for this picture comes from spectropolarimetry studies (Cohen
et al.\ 1995; Ogle et al.\ 1999; Brotherton et al.\ 2001). As a result,
BAL ionic column densities ($N_{ion}$) determined in the traditional
way (from equivalent width or direct conversion of the residual
intensity into optical depth) can only serve as lower limits to the
real column densities.  Since BAL column densities are the foundation
of any attempt to determine the IEA in the outflows, it became clear
that more sophisticated analyses are essential for any progress in
understanding the BAL phenomenon. To derive real $N_{ion}$ we have to
account for saturation and partial covering factor in the BALs.  In
the UV, this approach is currently feasible only for exceptionally
bright BALQSOs and still requires long integration times.  After
studying the available BALQSO data in the HST archive and hundreds of
ground based spectra, we concluded that PG~0946+301 is the best
candidate for such analysis. It has a substantially higher flux
between 1200--2500 \AA\ (observed frame) than any other BALQSO that is
suitable for IEA studies.  Thus, we can obtain very high-quality data
for a wide rest frame spectral region (500 -- 1700~\AA\/), and the
object suffers only minor contamination by \Ly\ forest lines due to
its low redshift ($z=1.223$). Finally, the BALs of PG~0946+301 are
typical for a high-ionization BALQSO. For example, the balnicity index
of the \civ\ trough is 6300 \kms, compared with 5000 \kms\ for the
BALQSO sample of Weymann et al.\ (1991).

It is for these reasons that many QSO researchers joined together
to study PG~0946+301 with deep, high-quality multi-wavelength
observations.  The components of the campaign were: HST UV
spectroscopy (described here), ASCA X-ray observations (Mathur et
al. 2000), FUSE UV spectroscopy (25 ksec data currently under analysis by
the FUSE PI team), high-resolution optical spectroscopy and optical
spectropolarimetry (some data already taken, other observations
planned).  

In this paper we describe the Space Telescope Imaging Spectrograph
(STIS) observations, present simple template fits for the BALs and
discuss our preliminary IEA findings.  We make the data (including
some ground based spectra) available to the community in order to
facilitate future analyses of this unique data set.  In \S~2 we
elaborate on the data acquisition, reduction and wavelength
calibration of the observations, and outline the temporal changes in
the BALs detected by comparing the 1992 HST/FOS spectrum  to the STIS
data. We defer most of the temporal analysis to a future
paper. In \S~3 we describe our template fitting analysis for the
entire spectrum, including the derivation of optical depth templates,
line identifications and unidentified absorption features, and also
discuss possible improvements over the template fitting technique.  In
\S~4 we describe the IEA constraints obtained from the current
analysis and discuss the over-abundance of phosphorus. 
In \S~5 we summarize our results.

\begin{figure}
\centerline{\psfig{file=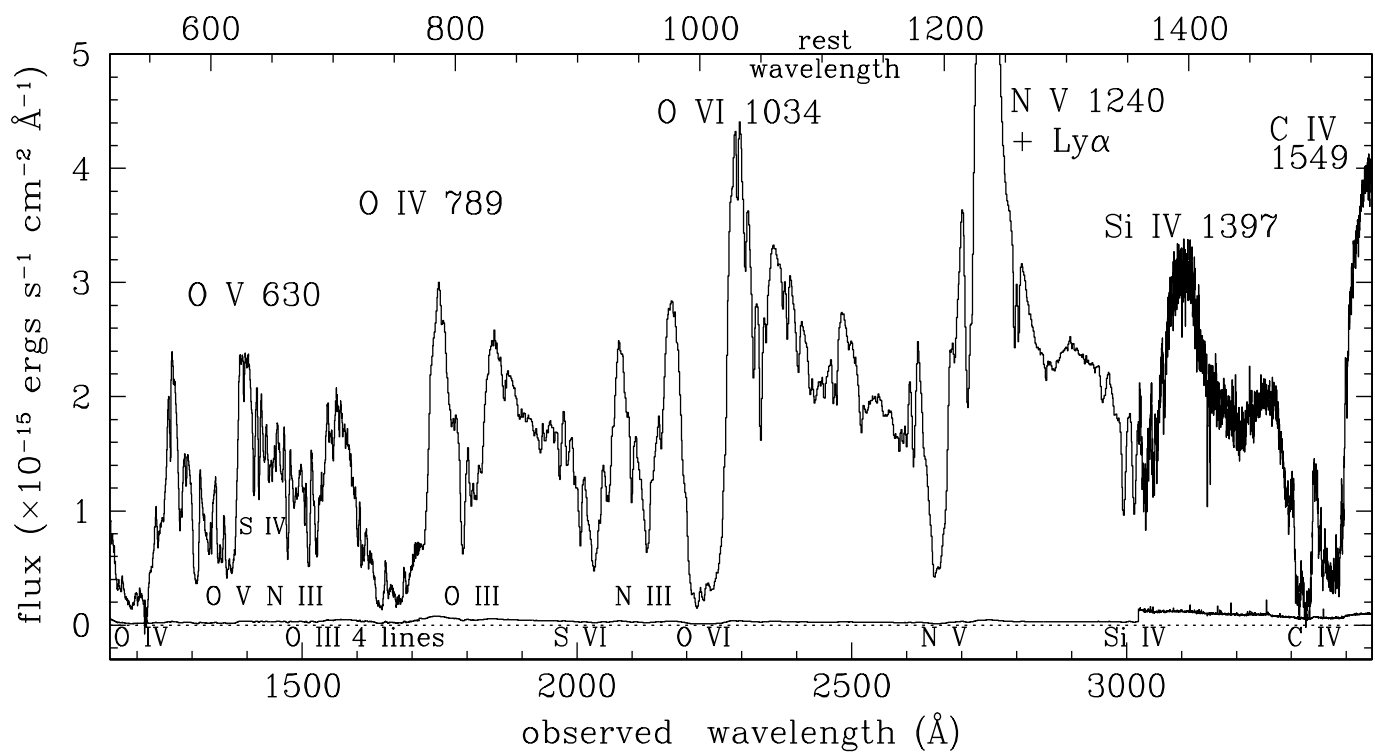,angle=-90,height=16.0cm,width=20.0cm}}
\vspace{-1cm}
\caption{Composite STIS spectrum of PG~0946+301. Flux is measured in
the observed frame.  The bottom labels identify the ions responsible
for the major absorption features.  Some of the broad emission lines
are marked above the spectrum.  The dotted line designates zero
intensity and the solid line above it is the noise
spectrum.}\label{fig1}
\end{figure}

\section{OBSERVATIONS}

\subsection{Data Acquisition and Reduction}

We observed PG~0946$+$301 with the Space Telescope Imaging
Spectrograph (STIS) on the {\it Hubble Space Telescope} ({\it HST})
over a three-month period in early 2000. All of the spectra were
obtained through a 52$''$ x 0\arcsecpoint2 slit to maximize throughput
without a significant loss in spectral resolution. We used the MAMA
detectors and G140L and G230L gratings to obtain full UV coverage at a
resolution of 1.2 \AA\ and 3.2 \AA, respectively. We also used the CCD
detectors and the G430M grating to obtain limited coverage in adjacent
optical regions at a resolution of 0.56 \AA. The details of the
observations are given in Table 1.

Each G140L or G230L entry in Table 1 represents a five-orbit visit, since STIS 
MAMA observations are limited to five consecutive orbits due to constraints 
imposed by the South Atlantic Anomaly. To maximize the signal-to-noise of the 
spectra, we obtained exposures at five different locations separated by 
0\arcsecpoint3 along the slit (one location per orbit). We used this strategy to 
place the spectra at slightly different positions on the detector, to avoid 
possible fixed pattern noise not removed by the flat-fielding of the spectra 
(Kaiser et al.\ 1998). For the G430M observations, we obtained two spectra per 
orbit (at the standard slit locations) to accumulate enough images to allow for 
accurate cosmic-ray rejection.

We reduced the spectra using the IDL software developed at NASA's
Goddard Space Flight Center for the STIS Instrument Definition Team
(Lindler 1999). The individual spectra from each orbit were
independently calibrated in wavelength and flux, and resampled to a
linear wavelength scale (retaining the same average
dispersion). Careful examination of the spectra from each orbit
reveals no discernible change in flux over time or as a function of
position on the detectors. We therefore averaged the spectra (weighted
by exposure time) to obtain final versions of the G140L, G230L, and
two G430M spectra. These were then combined into a single spectrum
using Galactic absorption lines for a final wavelength calibration.
The resultant spectrum and its noise are shown in Figure \ref{fig1}.

\begin{deluxetable}{ccrl}
\tablecolumns{4}
\footnotesize
\tablecaption{{\it HST}/STIS Observations of PG0946$+$031}
\tablewidth{0pt}
\tablehead{
\colhead{Grating} & \colhead{Range} & \colhead{Exposure} & \colhead{Date} \\
\colhead{} & \colhead{(\AA)} & \colhead{(sec)} & \colhead{(UT)}
}
\startdata
G140L &1140 -- 1715 &12,960 &2000 February 26 \\
G140L &1140 -- 1715 &12,960 &2000 March 28 \\
G140L &1140 -- 1715 &12,960 &2000 March 30 \\
G140L &1140 -- 1715 &12,960 &2000 April 4 \\
G140L &1140 -- 1715 &12,960 &2000 May 11 \\
G430M &3022 -- 3306 & 5,274 &2000 April 12 \\
G430M &3163 -- 3447 & 7,250 &2000 April 12 \\
G230L &1640 -- 3148 &12,960 &2000 April 23 \\
G230L &1640 -- 3148 &12,960 &2000 May 2
\enddata
\end{deluxetable}


\subsection{Comparison of the FOS and STIS Data}

\begin{figure}
\centerline{\psfig{file= 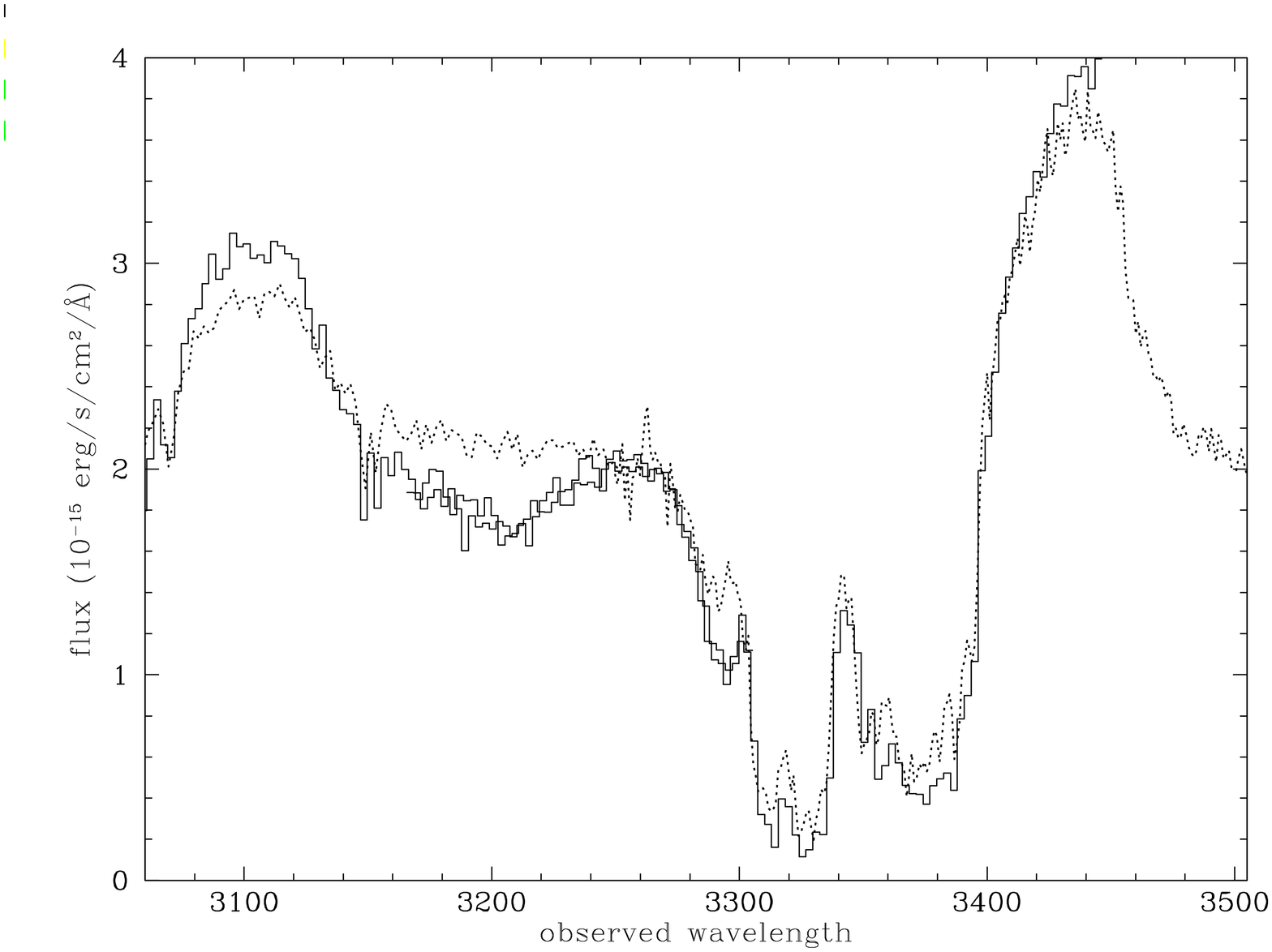,angle=-90,height=16.0cm,width=20.0cm}}
\vspace{-1cm}
\caption{ The \civ\ spectral region in PG~0946+301 at two different
epochs. Histogram: STIS data from March 2000; dotted line: FOS and Lick
data from 1992. The two settings of the G430M STIS grating are shown
(rebinned by 10 pixels), these overlap between 3165 -- 3300
\AA. In the main trough (3260 -- 3400 \AA) there is a significant
deepening of the absorption in the 2000 data and a new high velocity
trough is seen between 3130 -- 3250 \AA. We note that the differences
are very significant compared to the noise level and that the STIS
data are in remarkable agreement with Lick data from the same epoch
(not shown here).}
\label{variability}
\end{figure}

PG~0946+301 was observed by the HST Faint Object Spectrograph (FOS) in
1992. These data are still among the highest quality UV spectra of BALQSOs 
 (complete analysis of these observations is found in Arav et
al. 1999a, hereafter FOS99).  Combined with the much higher S/N
STIS data, we possess a unique data set for studying the temporal
changes in the many BALs observed in the UV.  In addition, we have five
epochs of the \civ\ trough obtained from ground based telescopes.  We
defer a full analysis of the observed temporal changes to a separate
paper.  Here we briefly describe the main changes and
concentrate on the ones that affect our model fits most.

In Figure \ref{variability} we show the \civ\ spectral region at two
epochs separated by 8 years ($\sim$3.5 years in the rest frame of the
object).  The 1992 spectrum is a combination of the HST data from the
FOS G270H grating ($\lambda<3250$ \AA, with a four pixel rebinning),
and Lick data ($\lambda>3250$ \AA).  The 2000 observations consist of
two setting of the STIS G430L grating rebinned by 10 pixels (see Table
1 for exact dates and exposure times).  Our Lick 2000 data (not shown
here) are in remarkable agreement with the STIS G430L data.
Inspection of Figure \ref{variability} reveals several significant
changes between the 1992 and 2000 spectra.

\vspace{-.2cm}
\begin{enumerate}
\itemsep=0pt
\parskip=0pt

\item Significant deepening of the main trough has occured between 3260 -- 3400
\AA\ (observed frame). For comparison the S/N level at the deepest
part of the trough is 6.5 for the Lick 1992 data and 7 for the binned STIS
data.

\item A substantial decrease in flux is seen between 3280 -- 3300
\AA\ ($\sim$ 25\%).  A similar change is also easily identified  in the
\nv\ and \ovi\ BALs.

\item A new shallow high velocity  trough is seen in the STIS data
between 3130 -- 3250 \AA\ (--17,200 to --28,400 \kms), with a maximum
depth of 20\%. Several other high ionization BALs show this feature
(most evident in \nv\ and \ovi). The high velocity trough is not present
in low ionization species (\siiv, \ciii, \niii...).

\item The \siiv\ absorption depth in the broad low velocity trough
 (--4000 to --7000 \kms) has increased significantly (not shown in the
 figure), and this can also be seen in other low-ionization species
 (e.g. \oiii, \niii).

 \end{enumerate}

Both independent settings of the G430L grating show the new high
velocity BAL as well as the decrease in flux that is seen
between 3280 -- 3300\AA. We also note that while the \siiv\ BEL is 20\%
stronger in the STIS data, the \civ\ BEL has not changed appreciably
(based on the Lick 2000 data).

\section{TEMPLATE FITTING ANALYSIS}

The simplest way of analyzing an absorption line spectrum containing
many broad overlapping lines with a complex velocity structure is the
template fitting approach (Korista et al.\ 1992; FOS99). This method
uses the apparent optical depth (defined as $\tau=-ln(I_r)$, where
$I_r$ is the residual intensity seen in the trough, for more details
see FOS99) profile of a given ionic transition as a template to fit
all other transitions, where only a simple multiplicative scaling is
allowed.  The fitting process finds the optimum scaling values which
minimize the difference between the data and a model spectrum based on
using scaled versions of the template for all plausible
transitions. We did this in two ways. First, we used standard $\chi^2$
statistics, i.e., minimized the square of the differences between the
model and the data, weighted by the square of the Poisson
error. However, because the S/N ratio is very high everywhere, the
deviations between the model and the observations are almost entirely
due to uncertainties regarding the effective continuum (defined below)
and covering factor effects. Therefore, we also fit the spectrum using
a constant error rather than the Poisson error at each wavelength
point. In practice using a constant error or the statistical error
changes the parameters derived from the fit by less than 0.1 dex so
that the exact minimization method used is irrelevant.

The fitted scaling factors are translated to ionic column densities
$(N_{ion})$ using standard techniques (see below).  Two assumptions
must hold in order that the fitted $N_{ion}$ be good approximations
for the real ones.  First, the absorbing material covers the emission
source completely and uniformly, and scattered-photons do not
contribute appreciably to the residual intensity in the trough (see
discussions in Korista et al.\ 1992; Arav 1997).  Second, the
ionization equilibrium does not vary as a function of velocity.  As we
will show, the first assumption is not a good approximation for the
BALs in \0946.  In this case we must treat the fitted $N_{ion}$ as
lower limits for the real ones (FOS99, Arav et al.\ 1999b, de~Kool et
al. 2001).

\subsection{Methodology}

Using the assumptions described above, the optical
depth at a given wavelength $\lambda$ due to a transition between
levels $l$ and $u$ of species $n$ can be expressed as
\begin{equation} \tau_{nlu}(\lambda) =
\frac{ \pi {e}^2 }{m_ec} \lambda_{lu} f_{lu} 
\frac{X_{nl}}{X_0} \frac{dN_0}{dv} \biggr|_{v=v(\lambda_{lu} \rightarrow
\lambda )}  \label{tau_wl} 
\end{equation}
Here $\lambda_{{lu}}$ and $f_{lu}$ are the rest wavelength and
oscillator strength of the transition, $\frac{X_{nl}}{X_0}$ the
density ratio between ions of species $n$ in energy state $l$ and some
reference species for which the distribution of column density over
velocity $\frac{dN_0}{dv}$ is known, and ${v(\lambda_{lu}
\rightarrow \lambda})$ is the velocity that shifts a line with rest wavelength
$\lambda_{lu}$ to the wavelength $\lambda$. The distribution of column
density over velocity $\frac{dN_0}{dv}$ is usually derived from the
optical depth in a transition corresponding to a line that is not
blended with other lines:
\begin{equation} 
\frac{dN_0}{dv} = \frac{{m_ec} }{ \pi\/{e}^2 }
\frac{1}{\lambda_{0} f_{0} } 
\vy{\tau}{0}(v) \label{dNdv}
\end{equation}
It can be derived directly if the transition is a singlet (e.g. de
Kool et al.\ 2001), or by a deconvolution if the transition is a
multiplet (e.g., FOS99). The line profile $\vy{\tau}{0}(v)$ of the reference
transition thus serves as a template for all other absorption lines.

Once the template is known, a model of the total optical depth as a
function of wavelength is constructed based on a first guess of the
ratio of the column density of each species relative to that of the
reference species, $\frac{X_{nl}}{X_0}$:
\begin{equation}
\tau (\lambda)= \sum_n \sum_l \sum_{u>l} \tau_{nlu} (\lambda) 
\end{equation}
Using the derived $\tau (\lambda)$ and a model for the unabsorbed
continuum, a model spectrum is computed that can be compared with the
observed spectrum. Using an iterative procedure, the values of
$\frac{X_{nl}}{X_0}$ are then adjusted until the $\chi^2$ is
minimized. The value of $\frac{X_{nl}}{X_0}$ of the best fit yields
the ratio of the column density of ion $n$ in energy state $l$ to the
column density of the reference species. In the current analysis we
assume that the ground-term level populations of the ions we consider
are in LTE, at a temperature much higher than the level energy, so
that the relative population of each energy level in the ground-term
is proportional to its statistical weight.

\subsection{The Velocity Templates}

\begin{figure}
\plotone{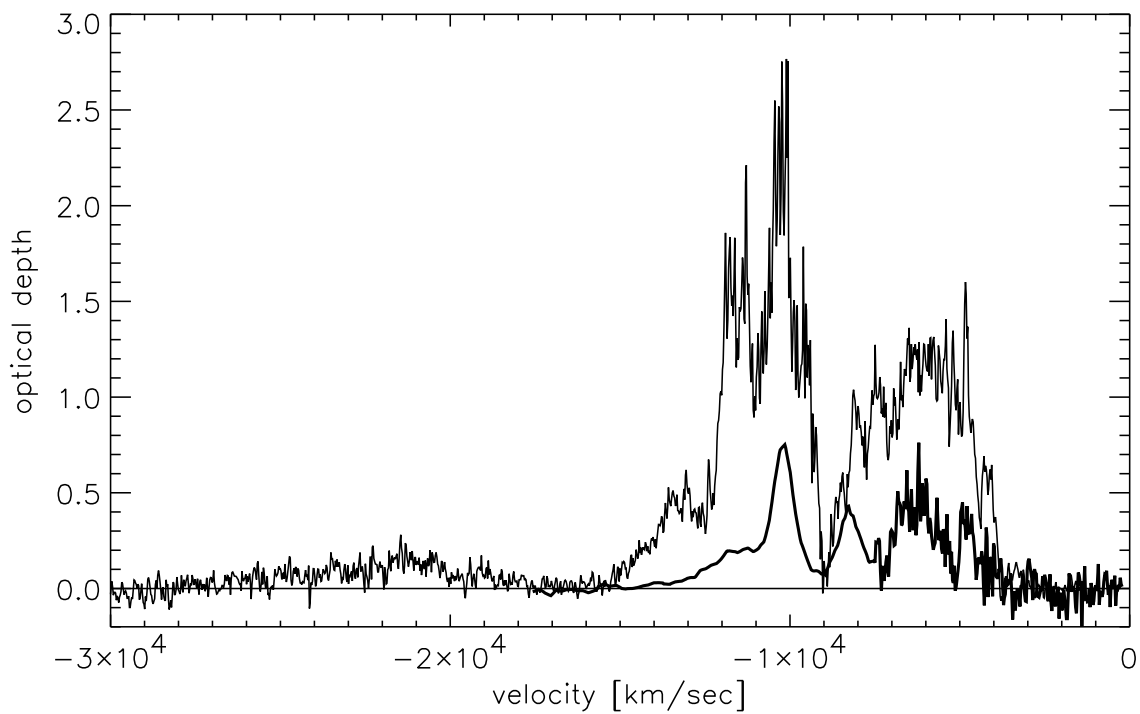}
\caption{The apparent optical depth as a function of velocity for the blue
components of the \civ~$\lambda$1549, (thin line) and
\siiv~$\lambda$1397 (thick line) resonance
doublets. }\label{templates}
\end{figure}

As in our previous study of PG~0946+301 we used templates derived from
two reference transitions. \civ~$\lambda$1548.20 is used for lines
from high ionization species, and \siiv~$\lambda$1393.76 for lines
from low ionization species, both lines are the blue component of
their respective doublets.  In Figure \ref{templates} we present the
\civ\ and \siiv\ optical depth as a function of velocity derived from
the STIS spectrum of PG~0946+301. The template for \civ\ was obtained
with a regularization method similar to that described in
FOS99. Because of the much higher resolution in the observed \civ\
line, the regularization used in FOS99 which suppressed fluctuations
in neighboring points did not give satisfactory results. Instead, we
used an alternative form of the regularization matrix {\bf \it H}, in
the notation of FOS99: $h_{i,i}=2$, $h_{i,i+s}=-1$, $h_{i,i-s}=-1$ and
all other elements of {\bf \it H} equal to zero. This form suppresses
oscillations on the scale of the doublet separation but does not
smooth the derived profile over neighboring points. The value of the
regularization parameter used was 0.1. We derived the \siiv\ template
 without regularization, since oscillations on the scale of
the doublet are not apparent in the plain deconvolution result.

The transition between data from the G230L grating and the G430M
grating, and the accompanying change in spectral resolution, occurs in
the middle of the \siiv\ line. This accounts for the smooth appearance
of the high velocity part of the \siiv\ template. There is no evidence
of the new high velocity trough in the \siiv\ absorption line. The
\siiv\ template was corrected for the presence of intervening \civ\
absorption at 1330 \AA\ in the QSO restframe by removing this line from
the spectrum before the deconvolution. We note that some of the
absorption in the range --11,000 to --13,000 \kms\ may be due to \civ\
from another intervening system that can be identified from its \Ly\
 absorption (component 9 in Table 2 in FOS99).
There is some
evidence for this being the case from the fact that the \siIV\ template
over-predicts absorption in other low ionization species (eg, \oiii,
\siv) at these velocities.

\subsection{The Effective Continuum}

In order to make measurements or fit the observed BALs we must use a
model for the unabsorbed emission (continuum plus BELs, hereafter
 ``the effective continuum'') of the quasar.  At wavelengths longer than
$\sim$ 1000 \AA, the spectrum of PG~0946+301 is very close to the
non-BALQSO composite of Weymann et al.\ (1991), which we therefore used
as our long wavelength effective continuum with only small
modifications to some emission line strengths (as in FOS99).
Shortwards of 830 \AA\ we found only two narrow regions (624 -- 631 \AA\ and
567 -- 571 \AA)  free from BAL features.  This complicates
the task of determining the effective continuum based on the data at
hand.  Unfortunately, very little is known observationally about
intrinsic QSO spectra in the range 500 -- 900 \AA. The QSO composite
spectrum of Zheng et al.\ (1998) does cover these wavelengths, but very
few QSOs contribute to the composite in this wavelength
range. Furthermore, the signal to noise ratio is so low that the use
of this composite as a continuum would introduce too much noise to
the analysis.

With these uncertainties in mind, we used a similar approach to the one
we employed in the analysis of the HST FOS data.  Three effective
continua are employed (shown in Fig.~4). The first is a fixed continuum
which is very similar to the fixed continuum in FOS99 (for details see
\S~5.2 of FOS99). We lowered the flat continuum shortwards of 900 \AA\
by 5\% and added two BEL features, one for \ov~$\lambda$630 and one
for \nev~$\lambda$571.  Both BELs are required by features evident in
the STIS data (the quality of the FOS G160L data was too low to
justify the addition of these BELs).  Emission features of roughly the
same shape and wavelengths  are also seen in a smoothed
version of the quasar composite spectrum produced by Zheng et
al. (1998).

\begin{figure}
\centerline{\psfig{file=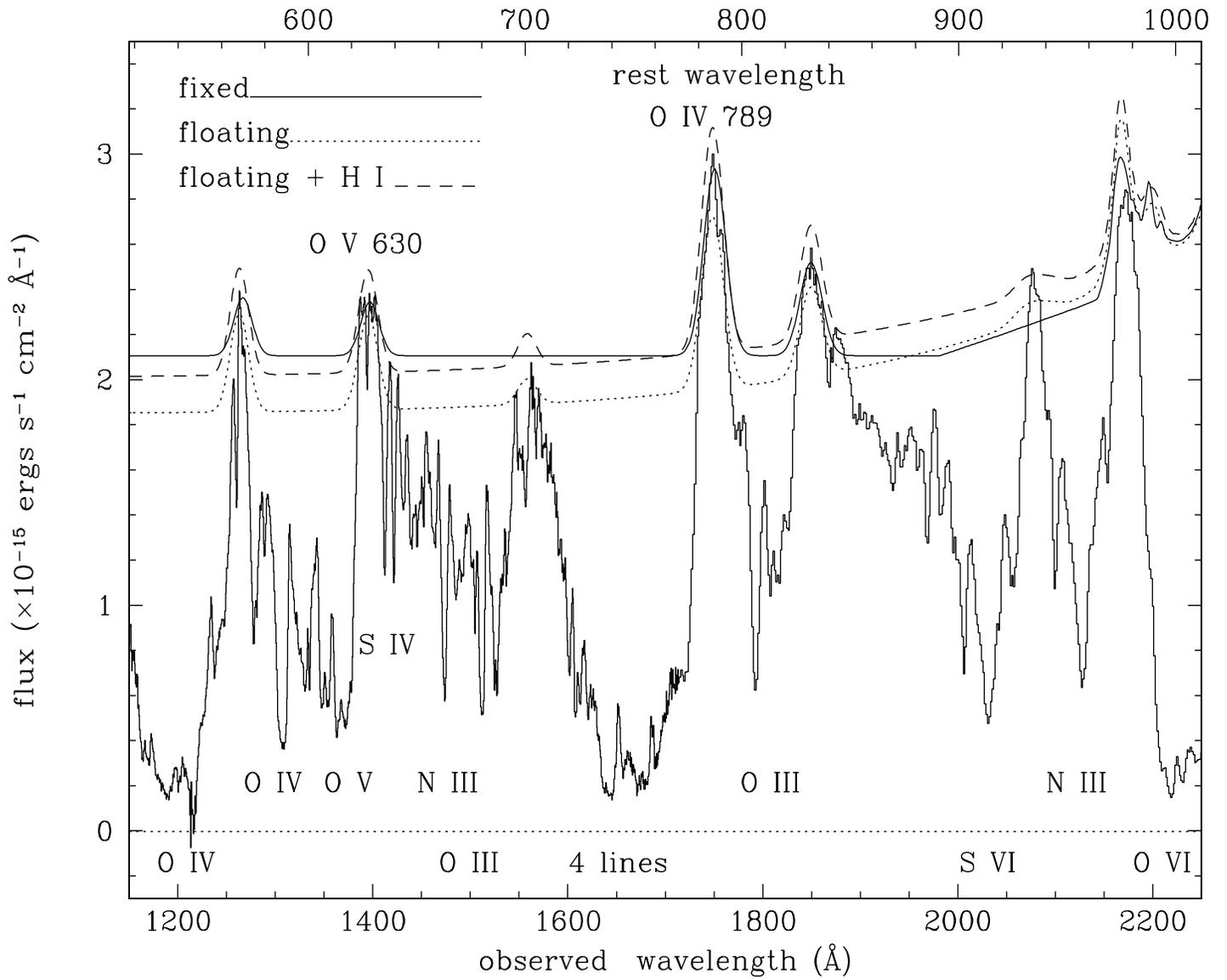,angle=-90,height=16.0cm,width=20.0cm}}
\caption{The three effective continua plotted on top of the data.
Longward of 1020 \AA\ they are indistinguishable from the one shown in Figure 
\ref{mainplot}. 
 }\label{fig_eff_cont}
\end{figure}

The second and third effective continua are different from the fixed continuum
for wavelengths below 1060 \AA. In this range, the underlying continuum flux
$F_c$ was assumed to have the functional form 
\begin{equation}
F_c =  F_0 + F_1 \bigl({ {\lambda}\over{1060}} \bigr)^{\beta}  
e^{-{{\lambda_{cut}}\over{\lambda}}}
\end{equation}
where $F_0$, $\beta$, $\lambda_{cut}$ and $F_1$ are free parameters
with the constraint that the continuum matches the long wavelength one
at 1060 \AA. The values of the continuum parameters were determined by
optimizing them simultaneously with the column densities of all
species ($\frac{X_{nl}}{X_0}$) during the template fitting
process. Since the continuum parameters were allowed to vary in the
fitting process, we refer to the resulting continuum as a floating
continuum. The difference between the two floating continua is that
one is derived from a template fit that only includes resonance lines
as opacity sources, whereas the other also allows for a Lyman edge
opacity (as a free parameter).  The equivalent widths of the model
BELs were fixed during the fitting process. Finally, most of the known
intervening absorption line systems were modeled with single Gaussians
with fixed equivalent widths to minimize possible confusion with BAL
structures.

\subsection{Results of the Template Fits}

\begin{figure}
\vspace{-1.5cm}
\centerline{\psfig{file=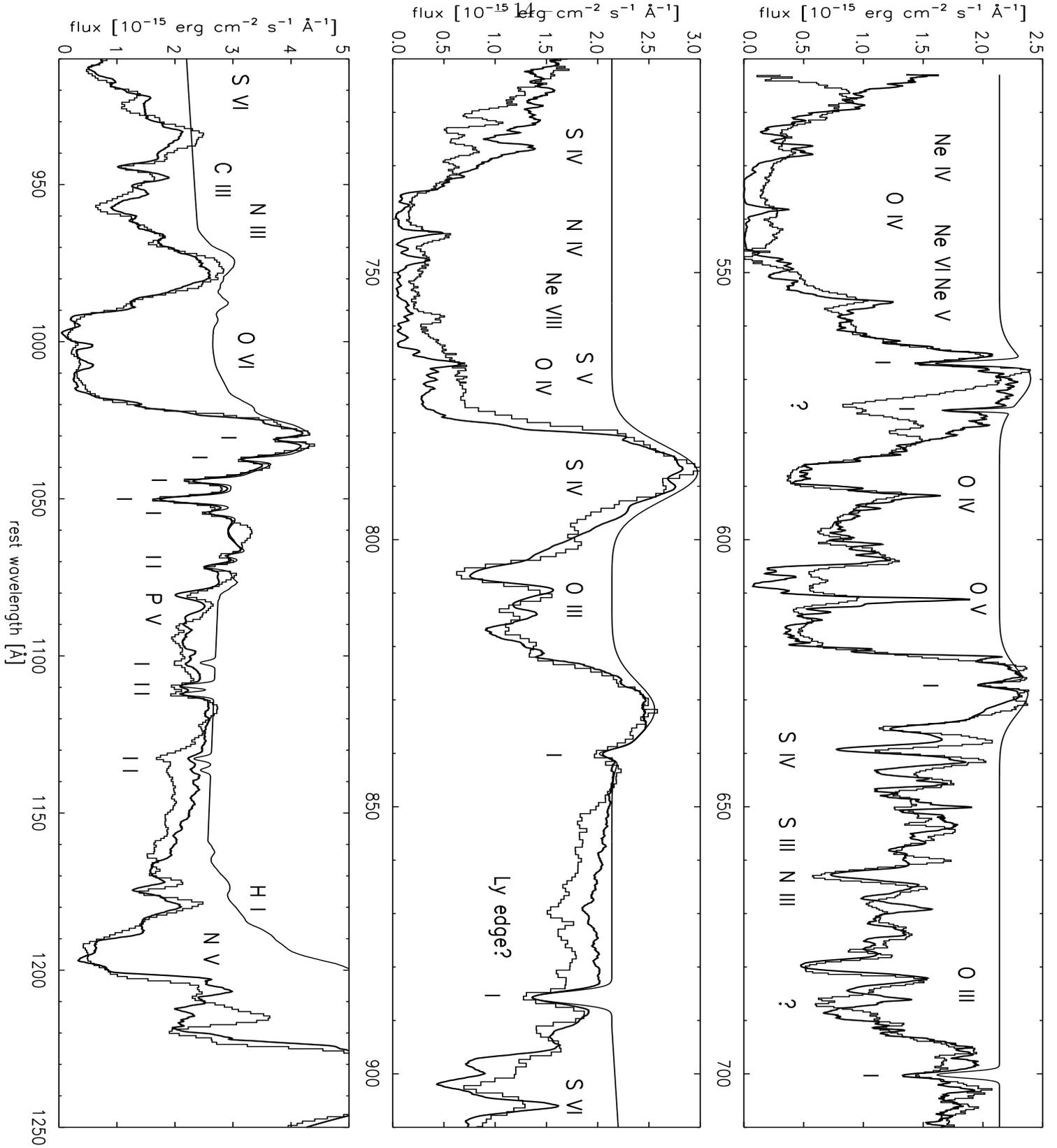,angle=0,height=22.0cm,width=20.0cm}}
\vspace{-.5cm}
\caption{Expanded view of a fit and the fixed effective continuum used in
producing it.  Most BAL features are identified by the absorbing
ion. The short vertical lines mark intervening and Galactic absorption
lines, which we superimpose on the effective continuum to account for
non-BAL absorption features.  
}\label{mainplot}
\end{figure}

The column densities resulting from the template fits are summarized
in Table \ref{col_table}, and the fit itself (for the fixed continuum)
is illustrated in Figure \ref{mainplot}. Before discussing these
results, we re-emphasize that column densities obtained from template
fitting are in reality only lower limits because of partial covering
effects. For some species we will also quote an upper limit on the
column density. Such an upper limit can be derived if the species in
question has at least one line with a much lower oscillator strength
than the others, which is not detected. These upper limits are
necessarily somewhat subjective since they depend on the shape of the
effective continuum and possible blending with other lines. By stating that
a line is not present we are in fact making a judgment as to what
variations in the effective continuum would be reasonable. Therefore, the
upper limits are derived in practice by manually changing the column
density in the fit and comparing the model spectrum and the observed
spectrum at the position of the weak line to see what column density
could be present without  conflicting with the observations.

When inspecting Table \ref{col_table}, the first thing to note is that
the derived column densities are insensitive to the exact choice of
effective continuum. Even changes of  30\% in the continuum level
affect the column densities of the stronger lines by a similar amount (i.e., 0.1 dex in the
table), while the effects of non-black saturation can increase the
inferred column densities by a factor of three or more.
 
Here follows a summary of the individual results:

{\it Hydrogen.} The \Lya\ BAL is not very deep, leading to a low value
for the column density ($\log(N_{\hi}) = 15.4$) derived from it. If
the fit includes a Lyman edge due to the BAL outflow, the best fit
Lyman edge corresponds to a column density of $\log(N_{\hi}) = 16.4$.
The neutral hydrogen column density from the Lyman edge fit is still
consistent with the upper limit derived from the absence of a clear
signature for Ly$\beta$ and Ly$\gamma$ BAL, $\log(N_{\hi}) < 16.7$. In
our ionization discussion (\S~4), we use a somewhat more conservative
upper limit of $\log(N_{\hi}) < 17$ to allow for partial covering
effects.

\begin{figure}
\centerline{\psfig{file=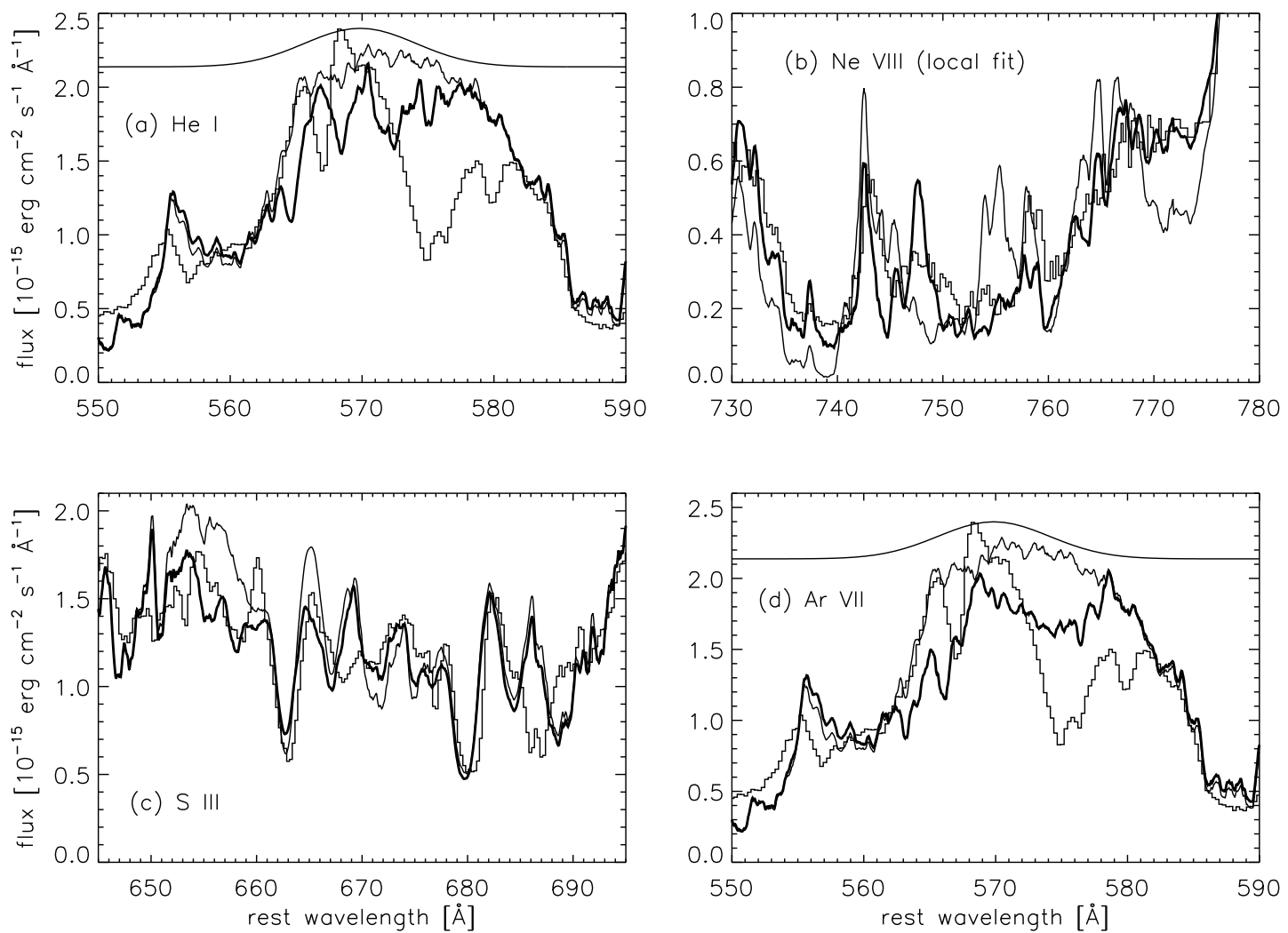,angle=90,height=16.0cm,width=20.0cm}}
\caption{Fit tests used to determine the existence of marginal, or
heavily blended species (see text for discussion). The observed
spectrum is shown as histograms, the thick and thin lines represent
models with and without the ion considered, respectively.
}\label{iffy_species}
\end{figure}

{\it Helium.} The upper limit quoted for \hei\ is derived from the
absence of a clear signature of the 584 \AA\ resonance line. This is
illustrated in Figure \ref{iffy_species}a.

{\it Carbon.}  We do not detect any absorption feature associated with
a \cii~$\lambda1335$ BAL.  The upper limit for this line provides
important ionization equilibrium constraints (see \S~4). A \ciii\ BAL
is clearly detected and its deepest component is not severely blended
with other BALs, so that the apparent column density is well
defined. The maximum optical depth is very similar to that of other
low ionization species (\hi, \niii, \siIV), which strongly suggests
that the low ionization BALs are essentially saturated and derive
their shapes primarily from partial covering. The \civ\ BAL is one of the
templates, so no new information can be derived from its template
fits.

{\it Nitrogen.} \niii\ gives rise to several lines in different parts
of the spectrum. The value for the \niii\ column density is driven
mainly by the strongest lines around 685 \AA, the other ones (991 \AA\
and 764 \AA) being both weaker and blended with stronger lines from
other species. The column density for \niv\ is not very well
determined since its BAL is a part of the large blended trough near
750 \AA. The \nv\ BAL is very deep, even though it falls under the
Ly$\alpha$ emission line.  This contrasts with the saturation effects
in the low ionization species mentioned above, which set in at a much
higher flux level and imply a low covering factor. The higher
ionization BALs appear to cover a significantly larger part of the
source than the low ionization BALs, including most of the Ly$\alpha$
BEL region..  This suggests a physical picture where at a given
velocity interval there is lower density extended material,
surrounding denser lower ionization material with smaller geometrical
extension.  (see FOS99 for a full discussion). The new shallow high
velocity component is deeper in \nv\ than expected based on scaling
the \civ\ template, as illustrated by the fact that the flux in the
fitted model significantly exceeds the observed flux at the
wavelengths where this absorption occurs. Note that this is not likely
to be a result of the uncertainty in the continuum level since a
comparison between the current STIS observation and the 1992 FOS
observation clearly shows the emergence of the new absorption
component. The fact that the high velocity component is deeper in \nv\
than in \civ\ suggests that it is highly ionized, which is supported
by the strength of this component in \ovi\ and \svi.

{\it Oxygen.} Oxygen is a unique element in this analysis since we
have relatively unblended BALs for four consecutive ionization
stages. \oiii\ has two multiplets at 703  \AA\ and 835 \AA. The \oiii\ BALs
are reasonably well fit by the model. The maximum optical depth in
\oiii\ is higher than in the other low ionization species, which is
consistent with the trend that ions with higher ionization potential
(IP) tend to have a larger covering factor (The \oiii\ IP is almost
midway between the \siiv\ and \civ\ IPs). Unfortunately, there are no
\oiii\ lines that allow us to determine an upper limit for the column
density.  \oiv\ has three groups of lines, and the fit is mainly
driven by the one with the lowest oscillator strength around 609
\AA. The other two BALs (554 \AA\ and 789 \AA) are part of the large
troughs around 550 \AA\ and 750 \AA. If the 609 \AA\ BAL is well fit,
the other two BALS are predicted to be deeper than observed, giving
rise to the very large $\tau_{max}$. Because this occurs close to the
bottom of the trough, the difference between model and fit does not
contribute as much to the overall $\chi^2$. Again, this demonstrates
the effects of partial covering. The \ov\ BAL is interesting in that
it clearly demonstrates how complex the real BAL formation is compared
to our simple model.  \ov~$\lambda630$ is a singlet and the main part
of the absorption (from --4000  \kms\ to --11,000 \kms) is not blended with
any other BALs. If our assumptions were correct, we would expect the
shape of the \ov\ BAL to correspond well to the \civ\ velocity
template. In reality, however the shape is quite different, with the
main velocity component in the template at --10,000 \kms\ being less
deep than the broad component between --4000 \kms\  and --9000 \kms. A
possible explanation for this behavior is that in this part of the
spectrum the source size is varying rapidly as a function of
wavelength, perhaps because of the presence of partially uncovered
emission lines. We consider it more likely that the problem lies with
continuum/source size problems around 600 \AA\ than with the template,
since many other high ionization lines, including \ovi, were fit very
well with the \civ\ template. \ovi\ also clearly shows the new shallow
high velocity component.

{\it Neon.} Absorption lines from several ionization stages of neon
are expected to occur in the spectrum of PG 0946+301, and the template
fitting procedure finds very large column densities for all of
them. However, the neon lines occur only within the two large blended
troughs around 550 \AA\ and 750 \AA\ so that their apparent column
densities are less reliable than those of most other ions. The
exception is \nev, its $\lambda$571 multiplet is the main source of
opacity for the red edge of the 555 \AA\ trough, and therefore its fitted
value is quite reliable.  The combination of the uncertainty in
the real optical depth in the neon lines with their low oscillator
strengths leads to large values for the column density that are
not necessarily realistic.  However, establishing the existence (or
lack) of a \neviii\ BAL is of particular importance in the context of
ionization equilibrium constraints (see \S~4).  We therefore conducted
a tailored fitting experiment to assess its existence.  We fit only
the 700 -- 845 \AA\ region and ignored the correlation with other lines
from the same ion found in different parts of the spectrum.  In this
way we circumvent some of the non-physical restrictions associated
with ignoring saturation and partial covering inherent in our fitting
process, and obtain the best local fit possible. Figure
\ref{iffy_species}b shows the data, a fit including \neviii\ and a fit
excluding it.  The large improvement caused by the inclusion of
\neviii\ is clearly seen in the 768 -- 777 \AA\, 753 -- 757 \AA\ and
730 -- 740 \AA\ regions.  We conclude that the presence of an
appreciable \neviii\ BAL ($\log(N_{\neviii})\gtorder16.5$) in \0946\
is highly probable.

{\it Magnesium.} In our analysis of the FOS data we discussed the
probable identification of the \mgx~$\lambda\lambda$~609,625 BAL, which
was based on fits to low-quality data obtained by the G160L grating.
Our vastly superior quality STIS G140L data do not show any features
that can be attributed to this BAL.  The excellent fit for the
580 -- 605 \AA\ region shown in Figure \ref{mainplot} does not include
any \mgx\ contribution. However, the strong presence of BALs
from \oiv\ and \ov\ do not allow the determination of an interesting upper
limit for this ion, the value we obtain for the upper limit is 
1--3 $\times10^{16}$ cm$^{-2}$.  

{\it Silicon.} We do not detect any absorption feature associated with
a \siII~$\lambda1263$ BAL.  The upper limit for this line supplies
similar ionization equilibrium constraints to those available from
\cii\ (see \S~4). A possible \siIII~$\lambda1206$ BAL would be blended
with the \Lya\ and \nv\ BALs, so that very tight constraints are not
possible. There is no evidence of the main component at --10,000 \kms\
in the spectrum, leading to the upper limit in Table \ref{col_table}.
Like \civ, \siIV\ is one of our template
species, therefore no new information is obtained from the fit.

{\it Phosphorus.} As in previous investigations, we find good evidence
for the presence of a \pv\ BAL. In spite of its high ionization
potential, the profile of the \pv\ BAL fits much better  with the
\siIV\ template than with the \civ\ template. This would be consistent
with a picture in which the BAL region contains a component with a very
high column density and a small covering factor, which is responsible
both for lines of lower ionization species and for lines from very low
abundance species (see FOS99 for a full discussion). A
possible \piv~$\lambda950$  BAL would occur in a highly blended region of
the spectrum, so that only a relatively uninteresting upper limit can
be derived.

{\it Sulphur.} \siii\ has a large number of lines that are predicted
to form very broad BALs without sharp features. Because of this,
identification of \siii\ features is not very secure, and \siii\ is
only marginally detected in the spectrum, as is illustrated in Figure
\ref{iffy_species}c. \siv\ is very clearly present, and the STIS
spectrum resolves the uncertainties associated with the \siv\ lines in
the FOS spectrum of PG~0946+301 (see FOS99). The depths of the
strongest \siv\ lines is similar to that of the low-ionization
template. \siv\ gives rise to several lines with a range of oscillator
strengths. This allows a determination of an upper limit on the \siv\
column density of $\log N(\siv) < 15.8$ (based on
\siv~$\lambda\lambda$809,815), which makes \siv\ the species with the
most tightly constrained column density in our analysis. \sv\ on the
other hand is very poorly constrained, and the global template fits
yield a negligible column density for it. The reason for this is that
the \sv\ line almost coincides with one of the stronger \oiv\ lines at
790 \AA. As mentioned above, the \oiv\ fit is driven by a very weak
line that already over-predicts \oiv\ absorption at the position of
the \sv\ BAL, so that the \sv\ column density is driven to zero. Since
\sv\ has only one line, this can occur without incurring penalties for
a bad fit in another part of the spectrum. The value quoted in Table 2
is derived from a local fit (700 -- 850 \AA), which suffers less from
the effects described above.  \svi\ is the strongest and deepest
sulphur BAL, in agreement with the hypothesis that higher ionization
species have a larger covering factor, although the BAL definitely has
a profile that is slightly different from that of the other high
ionization BALs. The shallow high velocity component is evident in
\svi, where it is deeper than that in \civ.

{\it Argon.} Resonance lines of \ariv, \arv, \arvi\ and \arvii\ occur
in the wavelength range covered by the STIS spectrum of
PG~0946+301. None of these lines are detected. For \ariv, \arv\ and
\arvi\ this leads to upper limits to their column density 
$\sim2\times 10^{16}$ cm$^{-2}$ because these species have lines with fairly
small oscillator strengths. These limits do not provide interesting
constraints. The situation is different for \arvii\ which has a strong
singlet (oscillator strength of 1.24) at 586 \AA. Figure
\ref{iffy_species}d shows a fit which includes the \arvii\ line at the
level of $1.3 \times 10^{15}$~cm$^{-2}$.  However, the fit is
unphysical since at the expected position of maximum \arvii\ optical
depth (567 \AA) we find a much higher residual intensity than the
fitted value.  (We note that the adjacent narrow absorption feature
seen in the data just longward of this position is due to Galactic
\siII~$\lambda$1260). In all the well identified BALs the deepest
predicted feature is fitted much better.  In addition the \arvii\ fit
does a very poor job in reproducing most of the surrounding spectral features.
Based on these arguments, $10^{15}$~cm$^{-2}$ is a conservative upper limit for 
the \arvii\ column density.

{\it Iron.} The spectrum was searched for the presence of BALs from
\feiii\ and \feiv.  \feiii\ has approximately 130 transitions in the
wavelength range considered, and if present, would give rise to
relatively broad blended BALs, similar to the case of \siii. The
strongest lines occur around 860 \AA, and the absence of any
detectable features corresponding to these leads to $\log N(\feiii)<
15.3$.  \feiv\ has a resonance multiplet centered around 525
\AA. Including \feiv\ does improve the fit at the very shortest
wavelengths, but there is no clear feature corresponding to it. 

{\it Unidentified absorption features.} We found several absorption features
in the STIS data for which we do not have good identification with either 
expected BALs, or with intervening and Galactic absorption lines.
Most prominent are the features at (observed wavelength):
1526 \AA\ ($\tau\sim$1),
1484 \AA\ ($\tau\sim$0.3),
1463 \AA\ ($\tau\sim$0.3),
1277 \AA\ ($\tau\sim$1). We note that the feature at 1526 \AA\ is too deep
and broad to be fully explained by Galactic \siII~$\lambda$1526 line, since it
should not be stronger than the observed \siII~$\lambda$1260 feature.

\begin{deluxetable}{lllcllrll}
\tablecaption{\sc Fitted Column densities} 
\tabletypesize{\scriptsize}
\tablehead{
\colhead{Ion}
&\colhead{Lines}\tablenotemark{a}
&\colhead{ $f_{ul}$}\tablenotemark{b}
&\colhead{Template}\tablenotemark{c}
&\colhead{ $N_{ion}$}\tablenotemark{d,e}
&\colhead{$\tau_{max}$}\tablenotemark{f}
&\colhead{$\lambda (\tau_{max})$}\tablenotemark{g}
&\colhead{ $N_{ion}$}\tablenotemark{d,h}
&\colhead{ $N_{ion}$}\tablenotemark{d,k}
}
\startdata
\hi          & 1215.67  & 0.416    & A  &    15.4 &   0.82 & 1215.67  &    -     &  -     \\
{\rm Ly lim} &  911.76  &   -      & A  &     -   &   -    & -        &    -     & 16.4   \\ 
\hei         &  584.33  & 0.285    & A  & $<$15.5 &   -    & -        &    -     &  -     \\ 
\cii         & 1335     & 0.127    & B  & $<$14.7 &    -   & -        &     -    &   -    \\ 
\ciii        &  977.02  & 0.767    & B  &    15.1 &   0.82 & 977.02   &    15.2  & 15.2   \\ 
\civ         & 1549     & 0.285    & A  &    16.1 &   3.34 & 1549     &    -     &  -     \\ 
\niii        &  991     & 0.122    & B  &    15.6 &   0.84 & 686      &    15.6  & 15.6   \\ 
             &  764     & 0.082 \\
             &  686     & 0.402 \\
\niv         &  764.15  & 0.616    & A  &   16.1: &   4.22 & 764.15   &    16.1: & 16.0:  \\ 
\nv          & 1240     & 0.235    & A  &    16.2 &   2.88 & 1240     &    -     &  -     \\ 
\oiii        &  834     & 0.107    & B  &    16.3 &   1.31 & 834      &    16.3  & 16.2   \\
             &  703     & 0.137 \\
\oiv         &  789     & 0.110    & A  &    16.9 &   8.83 & 554      &    16.8  & 16.8   \\ 
             &  609     & 0.067 \\
             &  554     & 0.335 \\
\ov          &  629.73  & 0.515    & A  &    16.2 &   3.30 & 629.73   &    16.2  & 16.2   \\ 
\ovi         & 1034     & 0.199    & A  &    16.6 &   5.04 & 1034     &    16.6  & 16.6   \\ 
\neiv        &  543     & 0.233    & A  &   16.6: &   2.68 & 543      &    16.6: & 16.6:  \\ 
\nev         &  571     & 0.092   & A  &   16.8  &   1.68 & 571      &    16.8  & 16.6   \\ 
\nevi        &  561     & 0.090   & A  &   16.8:: &   1.56 & 561      &    16.8:: & 16.8::  \\
\neviii      &  774     & 0.302    & A  &   16.5: &   2.36 & 774      &    16.5: & 16.5:  \\ 
\siII        & 1263     & 1.18     & B  & $<$14.2 &   -    & -        &    -     & -      \\ 
\siIII       & 1206.50  & 1.68     & B  & $<$14.3 &   -    & -        &    -     & -      \\ 
\siiv        & 1397     & 0.784    & B  &    15.1 &   0.86 & 1397     &    -     & -      \\ 
\piv         &  950.66  & 1.16     & B  & $<$15.0 &   -    & -        &    -     & -      \\ 
\pv          & 1121     & 0.701    & B  &    15.0 &   0.36 & 1121     &    -     & -      \\ 
\siii        &  680     & 1.64     & B  &  14.9:: &   0.41 &  680     &    14.6::& 14.6:: \\ 
\siv         &  814     & 0.103    & B  &    15.3 &   1.06 &  660     &    15.2  & 15.1   \\ 
             &  750     & 0.749 \\
             &  660     & 1.17  \\
\sv          &  786.47  & 1.42     & A  &    15.2\tablenotemark{m} &   0.70 &  786.47 & -  & - \\ 
\svi         &  937     & 0.665    & A  &    15.8 &   1.81 &  937     &    15.8  & 15.8   \\
\arvii       &  585.75  & 1.24     & A  & $<$15.1 &   -    &  -       &    -     & -      \\
\feiii       &  861     & 0.131    & B  & $<$15.3 &   -    &  -       &    -     & -      \\ 
\enddata
\pagebreak
\tablenotetext{a}{Lines that play a part in our analysis. Multiplets are rounded 
to the nearest \AA.}
\tablenotetext{b}{Total oscillator strength of the multiplet.}
\tablenotetext{c}{A refers to the \civ\ template, B to the \siiv\ template.}
\tablenotetext{d}{log($N_{ion}$) in units of cm$^{-2}$.}
\tablenotetext{e}{Fixed continuum.}
\tablenotetext{f}{The maximum optical depth due to the ion.}
\tablenotetext{g}{Wavelength (\AA) of the multiplet for which $\tau_{max}$ occurs.}
\tablenotetext{h}{Floating continuum (lines below 1060 \AA\ only).}
\tablenotetext{k}{Floating continuum including Ly edge (lines below 1060 \AA\ only).}
\tablenotetext{m}{The \sv\ BAL is strongly blended with 
\neviii, \niv\ and \oiv\ BALs. Therefore, we only used a regional fit (700--845 \AA) 
with the fixed continuum to constrain its column density.}
\tablenotetext{}{(:) spectral features from this species definitely present, but column density
uncertain. (::) spectral features from this species only marginally detected. }
\label{col_table}
\end{deluxetable}

\clearpage

\subsection{Beyond Template Fitting: Deriving and Constraining Real
   $N_{ion}$}

As discussed in the introduction, a main goal for further analysis of
the \0946\ STIS data is to derive real $N_{ion}$ for the BALs
or at least useful constraints beyond the simple template lower limits.
While not attempting such a full analysis here, we outline several
promising approaches as well as potential difficulties for doing so.

An obvious place to start is to examine the departure of the template fits
from the data.  In cases where we have BALs from two or more lines from
the same ion, these departures often  indicate non-black saturation and
may be used to determine the covering factor of the absorber. Two good
examples are the \oiv\ lines $\lambda$609 and $\lambda$554 with an
oscillator strength ratio of 1:5.  Inspection of our fits
clearly shows that the $\lambda$554 line is expected to be black if the
$\lambda$609 is well fit.  Since the $\lambda$554 trough is not
black, we conclude that this is a case of non-black saturation where
the covering factor can probably be measured directly from the shape
of the $\lambda$554 trough. A  similar case is found for \niii\
$\lambda$991 and $\lambda$686.  In order to estimate the degree of
saturation, a comparison can be made between global fits for the
\niii\ $\lambda$991 and $\lambda$686 lines and fits that assume these
lines are independent.  As a refinement one can try fits where
individual multiplet components are saturated.
Probably the easiest $N_{ion}$ to measure is that of \siv.  We
detect four \siv\ multiplets  at 1070 \AA, 814 \AA, 750 \AA\ and 660 \AA, with
oscillator strength ratio of 1:2:15:24.  Modeling of the
troughs associated with these multiplets should yield an even more accurate
$N_{ion}$ for \siv\ than we have currently.
 
A large volume of work (e.g., Arav 1997; Barlow 1997; FOS99; Churchill
et al.\ 1999; Arav et al.\ 1999b; de~Kool et al.\ 2001) shows that both
the level of saturation and the covering factor can be strongly
dependent upon the velocity.  Therefore, the extraction of $N_{ion}$
and the following determination of the IEA in the flow should be done for
as many velocity components as possible.  Ideally, it would be best to
work with the column density as a function of velocity instead of the
velocity integrated column density.

The task of extracting $N_{ion}$ is complicated by the following
difficulties: 1) Most doublet and multiplet components are blended
together since the velocity width of the BAL is much larger than the
separation between the different components of the same line.  If the
source is completely covered and there is no scattered light component
at the bottom of the troughs, an optical depth solution can be
extracted from blended components (Junkkarinen, Burbidge \& Smith
1983, FOS99).  However, this is obviously not the case for most BALs
present in the spectrum of \0946.  2) The considerable width of the
BALs causes blending between different lines.  The most extreme
example is the deep trough between 720 -- 770 \AA, which is a blend of at
least four strong BALs and probably a few additional weaker ones.  3)
The new high-velocity trough associated with the high ionization
species (\civ, \nv, \ovi...) worsens the BAL-blending
problem. Although not very deep, the high-velocity trough extends over
a velocity width comparable with that of the main deep trough.
Therefore, the combined velocity width of the BAL system in a given
high ionization line has doubled compared with the situation in 1999,
which worsens the BAL blending.

\section{CONSTRAINTS ON THE IEA IN \0946}

The template fitting results yield important constraints for the
ionization equilibrium and abundances (IEA) in \0946. All the
available  lower limits on $N_{ion}$, as well as upper limits based on non 
detections, can be used to constrain the IEA. The following findings
are particularly useful in doing so:

\vspace{-.2cm}
\begin{enumerate}
\itemsep=0pt
\parskip=0pt

\item There is no clear signature for either Ly$\beta$ and Ly$\gamma$
BALs, or a BAL Lyman-limit, which yield the upper limit $N_{\hi}
\ltorder 10^{17}$ cm$^{-2}$. Our template fit for the \Ly\ BAL gives
the lower limit of $N_{\hi}\gtorder 3\times10^{15}$ cm$^{-2}$.

\item Lower limits for the ionization parameter can be derived from the 
absence of neutral helium (inferred from the non detection of a
\hei~$\lambda$584 BAL) and singly ionized metals (\cii, \nii, \siII...)

\item A departure from solar abundances ratio is implied by the presence of
\pv\ combined with the absence of \arvii, the \hi\ upper limit and the
\siv\ constraints.

 \end{enumerate}

In this paper we concentrate on constraining simple-slab
photoionization models (using the photoionization code {\sc CLOUDY};
Ferland 1996).  Such models are commonly used in the study of quasar
outflows (e.g., Weymann, Turnshek, \& Christiansen 1985; Arav \& Li
1994 Hamann 1996; Crenshaw \& Kraemer 1999; de~Kool et al.\ 2001) and
assume that the absorber consists of a constant density slab
irradiated by an ionizing continuum.  The two main parameters in these
models are the thickness of the slab as measured by the total hydrogen
column density ($\vy{N}{H}$) and the ionization parameter $U$ (defined
as the ratio of number densities between hydrogen ionizing photons and
hydrogen in all forms).  The spectral shape of the ionizing continuum
is also important.  For this analysis we used two specific spectra:
The first is the standard Mathews--Ferland (MF) AGN spectrum (Mathews \& Ferland
1987).  The MF spectrum is very hard between 1--4 Ry, on the average
$f_{\nu}\propto\nu^{-0.8}$ over that range (the so called big blue
bump).  In \0946\ we do not see such spectral
characteristics. Instead, we observe $f_{\nu}\propto\nu^{-2}$ between
1--1.5 Ry with a hint of an even softer slope thereafter (see
Fig.~\ref{fig_eff_cont}). We therefore constructed a more realistic
ionizing-continuum model for \0946.  Below 1 Ry and above 40 Ry we
match the spectral shape (and relative intensities) of the MF
spectrum.  between 1--40 Ry we use an $f_{\nu}\propto\nu^{-2}$
slope. This shape is consistent with the observed spectrum while
keeping the shape of the far UV -- soft X-ray spectral region as
simple as possible.  We note that the expected Galactic extinction in
the direction of \0946\ is negligible (E[B-V] = 0.018 mags., Schlegel,
Finkbeiner \& Davis 1998), and therefore should not cause an
appreciable correction for the intrinsic UV spectral slope.  The
results described below were derived using this latter spectral shape.
For comparison we discuss the quantitative differences which arise
from using the MF spectrum at the end of this section.

\begin{figure}
\centerline{\psfig{file=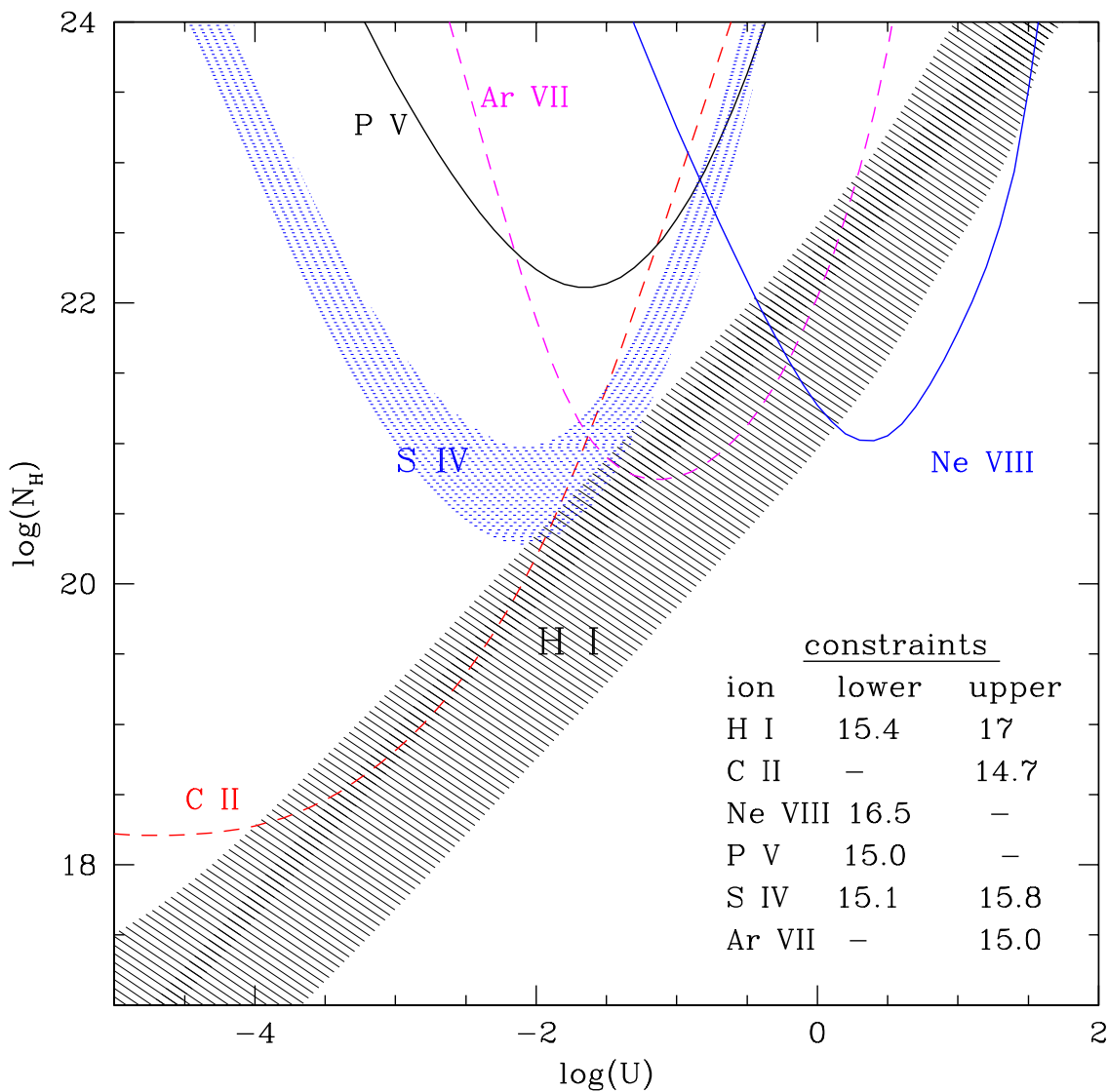,angle=0,height=16.0cm,width=16.0cm}}
\vspace{-1cm}
\caption{Curves of constant ionic column density plotted on the plane
of total hydrogen column density ($\vy{N}{H}$) of the slab vs. the
ionization parameter of the incident radiation ($U$), using solar
abundances.  Lower limits are shown as solid lines, upper limits as
dashed lines, shaded regions designate cases where we have both limits
for a given ion. The inserted Table gives the $\log(N_{ion})$ for the
various constraints.}
\label{nh_u}
\end{figure}

A particularly useful way to constrain such models is to plot curves
of constant $N_{ion}$ on the plane of $\vy{N}{H}$ vs. $U$.  On such a
plot any type of information regarding the ionic column densities can
be used to constrain the allowed parameter space.  Lower limits
(derived from fits and/or apparent optical depth measurements) exclude
the area below the constant $N_{ion}$ curve, $N_{ion}$ upper-limits
(from non-detections) exclude the area above their associated
constant-$N_{ion}$ curve.  In Figure \ref{nh_u} we show the most
significant $N_{ion}$ constraints extracted from the STIS data where
we use the $f_{\nu}\propto\nu^{-2}$ spectrum described above as the
incident ionizing continuum and assume solar abundances. For \hi\ we
have both a lower limit from our fit to the \Ly\ BAL and an upper
limit from the absence of clear signature for either Ly$\beta$ and
Ly$\gamma$ BALs, or a BAL Lyman-limit.  In our $\vy{N}{H}$ vs. $U$
presentation, the combined constraints restrict the allowed parameter
space to the diagonal line-shaded band.  A similar situation arises
for \siv\ where we derive a lower limit on $N_{\siv}$ from the strong
multiplet at 660 \AA\ and an upper limit from the weaker multiplets at
1070 \AA\ and 814 \AA.  The combination of both constraints gives the
dot-shaded area.

Several important conclusions can be drawn from Figure \ref{nh_u}.
The combined constraints from \hi, \siv, \cii, \arvii\ as well as from
other available CNO ions (which for clarity's sake are not shown
here), suggest that the absorber is characterized by
$20.3<\log(\vy{N}{H})<20.8$ and $-1.7<\log(U)<-1.2$. This narrow range
in the $\vy{N}{H}/U$ plane is fully consistent with 23 out of our 27
ionic column-density constrains, without assuming  departure from
solar abundances.  However, the limits on the allowed metalicity range
are rather wide, reflecting the uncertainty in the \hi\ BAL column
density (see \S~3.4).  We find that a simple scaling of the metalicity
in our models allows for solutions consistent with the \hi\ constraints
for $0.7\Zsun<Z_{0946}<10\Zsun$.  The total column density of the
absorber is inversely proportional to its metalicity value.
 
There are two constraints which are inconsistent with this narrow
range in the $\vy{N}{H}/U$ plane.  The firm detection of \pv\ (and
even more so in the FOS data, see Junkkarinen et al.\ 1997) yields the
associated lower limit on Figure \ref{nh_u}, which is incompatible
with the allowed parameter space we derived above.  The simplest way
to resolve this apparent contradiction is to assume that phosphorus is
over-abundant by a factor of 10--30 compared to its solar
abundance. All else equal, this will lower the \pv\ curve to the point
where it intersects with the allowed parameter space. Further
discussion of the inferred phosphorus over-abundance is found in
\S~4.1 and in the Appendix.

Another inconsistency is evident from the \neviii\ curve, for which
the allowed parameter space is strongly excluded by most other metal
constraints.  Here there are two possibilities.  If the \neviii\
measurement is robust, our naive assumption of a single $U$ value (or
a narrow range of $U$) must be wrong.  A second option is to question
the \neviii\ detection since the \neviii\ BAL resides in the middle of
the large trough and is blended with strong \oiv, \sv\ and \niv\ BALs,
as well as from a few weaker ones.  However, as we show in \S~3.4 and
in Figure \ref{iffy_species} it is very probable that the wide
absorption trough around 750 \AA\ contains a considerable \neviii\ BAL
contribution. The constraints from \svi\ and \nev\ are also marginally
above the allowed area in the  $\vy{N}{H}/U$ plane (by $\sim0.2$
dex).  Since both ions are of relatively high ionization potential,
these results support the hypothesis of a larger range of higher $U$
values in the outflow, which is needed to explain the inferred
\neviii\ value. We also point out that the uncertainties in the
continuum's shape increase at higher energies (since our last
data point lies near 1.8 Ry).  Therefore, in the vicinity of these
high ionization potentials the extrapolated flux level is rather
uncertain, which may contribute to the inconsistency between
the model results and the observed constraints for these ions.

How relevant are these constraints to the actual \0946\ outflow?  Our
analysis of the FOS data (FOS99) suggests a range in the ionization
parameter at a given velocity.  Furthermore, it shows that
the velocity dependence of the apparent optical depth is different for
low ionization BALs than for high ionization ones (some of this is
accounted for by using different templates for high vs.\ low ionization
species).  We argue that these effects can only modestly change our
overall IEA conclusions, but there might be some minor outflow
components with differing constraints which are missed by ignoring the
velocity dependencies.  Several arguments support this assertion.  The
\pv\ relative ionic fraction is at a maximum inside the $U$ range already
derived from the other metals/hydrogen constraints, $-1.7<\log(U)<-1.2$.
Therefore, a wide range in $U$ will only increase the inferred
over-abundance of phosphorus.  More generally, inspection of the \hi\     
and \siv\ constraints shows that a similar conclusion applies for most
metals.  It is difficult to construct a multi $U$-zone model which will
reproduce a better \siv\ to \hi\ ratio without invoking a larger
departure from solar metalicity for both higher and lower $U$ models.

For comparison, using the MF spectrum for the same analysis presented
in Figure \ref{nh_u} leads to the following conclusions: The absorber
is characterized by $19.8<\log(\vy{N}{H})<20.3$ and $-2<\log(U)<-1.5$
and a minimum metalicity enhancement  of 2-3 compared to
solar. The metalicity enhancement is necessary since for the MF
spectrum the strips allowed by the \siv\ and \hi\ constraints do not
overlap, and most of the metals' lower limit curves lie somewhat above
the allowed \hi\ zone.

\subsection{The Phosphorus Over-abundance}

Previous claims for a large phosphorus over-abundance in \0946\ are found in
Junkkarinen et al.\ (1997). By comparing the derived column densities
of \civ\ and \pv\ and accounting for possible ionization differences,
these authors derived (P/C)$_{0946}\sim60$(P/C)$_{\odot}$.  However,
this assertion is crucially dependent on the assumption that the
column densities deduced from converting the residual intensities to
optical depth are a good approximation for the real ones.  Even at the
time, the possibility of saturation and covering factor was addressed
and Junkkarinen et al.\ acknowledged that: ``If the \civ\ column
density is underestimated by a large factor and the \pv\ column
density is underestimated by a small factor, then the apparent
over-abundance of phosphorus could be much less.''  Since there is now
ample evidence for saturation and varying covering factor in the BALs
of \0946, the phosphorus over-abundance claim based on the \pv/\civ\
ratio cannot stand on its own.

Our claim for a phosphorus over-abundance does not suffer from this
weakness since we do not treat apparent $N_{ion}$ (i.e., the template
fitting values) as real $N_{ion}$.  Instead we use $N_{ion}$ lower and
upper limits, which are largely immune to saturation and covering
factor effects.  As can be seen in Figure \ref{nh_u} a comparison
between the \pv\ lower limit and the \hi\ upper limit necessitates a
large phosphorus over-abundance in order for the two constraints to
overlap in parameter space.  The exact value will depend somewhat on
the shape of the ionizing continuum and on the spread of $U$ values,
but a phosphorus over-abundance by a factor of less than 10 relative
to solar requires a rather contrived photoionization model.  
We also checked whether optically thick (bound-free)
photoionization models could ease or eliminate the need for phosphorus
over-abundance.  Our conclusion is that these models do not improve on
the optically thin results, where the details of this investigation
are found in the Appendix. 

Finally, it must be noted that there remain major uncertainties in some
of the atomic physics implemented in spectral simulation codes such as
{\sc Cloudy}. Among these are many of the dielectronic recombination
coefficients, especially for 3rd and 4th row elements, such as P, S,
and Ar (see Ferland et al.\ 1998). Until more reliable coefficients
become known, any conclusions drawn regarding gas abundances of these
elements should be done so with caution.

\section{SUMMARY}

We have presented an HST/STIS spectrum of \0946, which constitutes the
 highest quality UV data set of any BALQSO. These data are both high signal
to noise and covers a very large number of BALs between 500~\AA\/ and 
1550~\AA\/ in the rest frame. This provides for much tighter
constraints on the ionization and abundances than previously possible. These
data are available to the community for further analyses at:\\
http://unix.cc.wmich.edu/korista/ftp/pg0946/spectra 

Our preliminary analysis of these data yield the following results:

\begin{itemize}

\item Most of the 27 BAL troughs detected are saturated while not
black (i.e., their shape is largely determined by partial covering
effects).

\item The main outflow component is characterized by a total column
density $20.3<\log(\vy{N}{H})<20.8$, ionization parameter
$-1.7<\log(U)<-1.2$ and is consistent with solar abundances.
However, the metalicity can be up to ten times solar due to the 
large uncertainty in the \hi\ BAL column density.
These findings take into account the complications posed by
the saturation mentioned above.

\item The abundance ratio of phosphorus to other metals is roughly ten times
the solar ratio.  Ionization models including a \heii\ ionization
front cannot significantly decrease the necessary phosphorus
over-abundance.

\item In addition to the main outflow component, a much higher
ionization component appears to be present. This wide range in
ionization parameter is inferred from the detection of \neviii\ and
supported by the constraints from other high ionization species.

\item A new shallow high velocity trough appeared in these spectra
that was not present in the 1992 HST/FOS spectra. This trough is
evident only in the higher ionization lines, and the depth of this
trough appears to increase with ionization potential of the
ion. Significant changes in depth occurred in the main trough as well.

\end{itemize}

\section*{ACKNOWLEDGMENTS}

Support for this work was provided by NASA through grant number
G0-08284 from the Space Telescope Science Institute, which is
operated by the Association of Universities for Research in Astronomy,
Inc., under NASA contract NAS5-26555. NA expresses gratitude for the
hospitality of the Astronomy depeartment at the University of
California Berkeley for the duration of this work.

\section*{APPENDIX: CAN MODELS WITH IONIZATION FRONTS LESSEN THE
REQUIRED PHOSPORUS OVER-ABUNDANCE?} 

Due to the significance of a phosphorus over-abundance, we made
efforts to check whether optically thick (bound-free) ionization
models can explain the observed $N_{ion}$ constraints without invoking
departure from solar metalicity ratios.  The most promising set of
such models are those that include a strong \heii\ ionization front
(Voit, Weymann \& Korista 1993), where the incident spectrum is
strongly suppressed beyond 54 eV, the ionization potential (IP) of
\heii. Such an ionizing continuum may protect \pv\ from
photo-destruction (IP=65 eV), while still allowing for the
photoionization of \hi\ (IP=13.6 eV) and \siiv\ (IP=47 eV).  In
principle this can narrow the gap between the \hi\ and \pv\ curves.
Another advantage of such a model is that it might explain the sharp
dichotomy we observe between the low ionization species (\siiv, \niii,
\siv\ and \hi) and the higher ionization species (\civ, \nv, \oiv,
\svi...), since the \heii\ ionization potential falls in between these
groups.  However, a possible problem is that \arvii\ should also be
protected by a \heii\ front (although to a lesser extent because of
its higher IP of 124 eV), but we do not detect it in the data, and due
to the high oscillator strength of the \arvii~$\lambda585.75$ singlet we
obtained a strong constraint on the upper limit of its column density
($<10^{15}$ cm$^{-2}$).

In order to test whether \heii\ ionization front models can eliminate
or lessen the need for phosphorus over-abundance, we constructed
numorous {\sc CLOUDY} models with such fronts.  Here we give a brief
summary of several representative models.
We attempted to find solar-abundances {\sc CLOUDY} models which
simultaneously reproduce the inferred upper limits to the column densities 
for \hi, \siv\ and \arvii, and the lower limit for the \pv\ column
density.  Due to the low solar-abundance of phosphorus, {\sc CLOUDY} models
tend to produce higher ratios of \hi/\pv, \siv/\pv\ and \arvii/\pv\
compared to the observed values, so that
the goal of our modelling is to bring the upper limits of these more
abundant species into agreement with the lower limits of \pv. The
observed column density ratio limits are quite constraining, since the
ionization potentials of the other ions bracket the IP of \pv\ (see
above).

We let {\sc CLOUDY} find the optimal model (i.e., finding a minimum
for the $\chi^2$ between the observed and model-produced constraints,
summed over the four constraints detailed above) by changing
$\vy{N}{H}$ and $U$.  The main difficulty is due to the unknown shape
of the incident ionizing continuum.  We experimented with many
continuum shapes and results from four representative models are shown
in Table 3: 1) The canonical Mathews and Ferland (MF) AGN spectrum. 2)
The $F_{\nu}\propto \nu^{-2}$ ionizing spectrum described in
\S~4. 3) A very steep spectrum of $F_{\nu}\propto \nu^{-3.5}$ above 1
Ry, which, although not realistic, comes closest to being consistent
with the observed column density ratio limits 4) A more physically
realistic version of the above, which takes into account the observed
$\nu^{-2}$ spectral slope and connects to the MF spectrum at high and
low energies.

As can be seen in the table, the column density ratios of \arvii\ and
\siv\ to \pv\ are 5--10 times higher than observed.  The main reason for
this is that if a slab of gas contains a \heii\ front there are at least
two ionization zones present: a zone with highly ionized gas in which He
is completely ionized (the \heii\ Stro\"mgren layer), and a second zone
behind it in which \heii\ becomes the dominant helium ion. The first
highly ionized zone contains a significant amount of \arvii. The spectrum
irradiating the second zone contains a \heii\ edge, and this zone contains
virtually all of the \pv.  Due to the high $U$ values of the models
(necessary in order not to over produce \hi) the first zone is so thick
that the \arvii\ column density constraint is already exceeded at a total
hydrogen column density at which no detectable \pv\ is predicted. If the
models try to lower $U$ in order to produce less \arvii\ before the front,
then too much \hi\ and \siv\ are produced.  The end result is a compromise
that keeps both \arvii/\pv\ and \siv/\pv\ ratio about an order of magnitude
above the observed values.  Only a very contrived ionizing spectral shape
can change this situation.  We note that if the outflowing gas sees an
ionizing spectrum which already includes a strong \heii\ front, it is
possible to satisfy all four observational constraints.  However, in that
case we should see evidence for the ``shield'' that created the \heii\
front in the spectrum.  Since such a ``shield'' is very probably static,
we should see deep absorption features corresponding to it in \civ, \nv\
and \ovi, close to the systemic velocity.  Such features are not detected
in the spectrum.

\begin{deluxetable}{lllrllccr}
\tablecaption{\sc photoionization models with \heii\ fronts} 
\tablehead{
\colhead{\#}
&\colhead{spectral shape}
&\colhead{$\log(\vy{N}{H}$)}\tablenotemark{e}
&\colhead{$\log(U)$}\tablenotemark{f}
&\colhead{$\vy{\tau}{\hi}$}\tablenotemark{g}
&\colhead{$\vy{\tau}{\heii}$}\tablenotemark{h}
&\colhead{$\frac{\mbox{\arvii}}{\mbox{\pv}}$}\tablenotemark{k}
&\colhead{$\frac{\mbox{\siv}}{\mbox{\pv}}$}\tablenotemark{l}
&\colhead{$\vy{N}{\hi}$}\tablenotemark{m}
}
\startdata
1\tablenotemark{a}  & MF                & 21.46  & -1.1  & 1.3 & 190 & 10 & 4  & $2\times 10^{17}$ \\
2\tablenotemark{b}  & $\alpha=-2$       & 21.67  & -0.85 & 0.9 & 460 & 10 & 9  & $1.4\times 10^{17}$ \\
3\tablenotemark{c}  & $\alpha=-3.5 $    & 21.54  & 0.00  & 0.1 & 330 &  4 & 4  & $1.4\times 10^{16}$ \\
4\tablenotemark{d}  & $\alpha=-2, -3.5$ & 21.52  & -0.76 & 0.5 & 290 &  6 & 6  & $8\times 10^{16}$ \\ 
\enddata
\tablenotetext{a}{Mathews and Ferland (MF) AGN spectrum, with a 10 $\micron$ break.}
\tablenotetext{b}{$F_{\nu}\propto \nu^{-2}$ for 1 Ry$<h\nu<$40 Ry and MF shape elsewhere.}
\tablenotetext{c}{$F_{\nu}\propto \nu^{-3.5}$ for 1 Ry$<h\nu$.}
\tablenotetext{d}{$F_{\nu}\propto \nu^{-2}$ for 1 Ry$<h\nu<1.5$ Ry, 
$F_{\nu}\propto \nu^{-3.5}$ for 1.5 Ry$<h\nu<30$ Ry and MF shape elsewhere.}
\tablenotetext{e}{Optimal $\vy{N}{H}$ for the slab in units of cm$^{-2}$.}
\tablenotetext{f}{Optimal $U$ value for the model.}
\tablenotetext{g}{Optical depth at the lyman edge.}
\tablenotetext{h}{Optical depth at the \heii\ lyman edge.}
\tablenotetext{k}{Ratio of modeled $\vy{N}{\mbox{\arvii}}/\vy{N}{\mbox{\pv}}$ normalized to the observed ratio.}
\tablenotetext{l}{Ratio of modeled $\vy{N}{\mbox{\siv}}/\vy{N}{\mbox{\pv}}$ normalized to the observed ratio.}
\tablenotetext{m}{Model value in units of cm$^{-2}$.}
\label{col_table}
\end{deluxetable}

\pagebreak

\end{document}